\documentclass[11pt]{article}
\usepackage[margin=1in]{geometry}
\usepackage{amsmath,amssymb,mathtools}
\usepackage{booktabs}
\usepackage{graphicx}
\usepackage{microtype}
\usepackage{tabularx}
\usepackage{array}
\usepackage{enumitem}
\usepackage{tikz}
\usepackage[hidelinks]{hyperref}
\usepackage{caption}
\usepackage{xspace}
\usepackage{multirow}
\usepackage{float}
\usetikzlibrary{arrows.meta,positioning,fit,calc,backgrounds}
\newcommand{\bsr}{Bounded-State Restoration\xspace}

\title{\textbf{Bounded-State Restoration:\\Decoupling Local Restore Capacity from External LLM State}}
\author{Zixuan Li\\
China Academy of Railway Sciences Corporation Limited, Beijing, China\\
\href{https://orcid.org/0009-0005-4713-3032}{ORCID: 0009-0005-4713-3032}}
\date{August 2026}
\begin{document}
\maketitle
\begin{abstract}
Hierarchical KV-cache systems can retain long-context LLM execution state far
beyond GPU memory, but retention capacity does not determine how much local
memory is needed to make that state executable again.  We isolate this second
resource as the \emph{restoration working set} (RWS): the peak local staging
state whose lifetimes overlap during restoration.  In the pinned upstream
LMCache whole-plan path, measured full-reuse points track the selected state:
1.956, 7.823, and 15.646\,GiB/rank states first succeed at 2, 8, and 16\,GiB
L1 rungs, and their successful L1 peaks are 1.956, 7.824, and
15.648\,GiB/rank.

We introduce \bsr (BSR), a protocol that separates complete discovery from
local residency.  BSR first probes the complete reusable prefix without
materializing the whole hit in L1, then installs confirmed state through a
reusable window of at most $W$ chunks.  Under bounded auxiliary state, peak
restoration capacity becomes $O(W)$ while total transfer and installation work
remains $\Theta(|S|)$.  Because reusable execution state spans heterogeneous
allocator groups and tensor-parallel ranks, BSR also uses a request-level commit
rule: partial installation is never exposed as a valid reusable prefix;
failures invalidate the advertised prefix and fall back to a lower valid tier
or deterministic recomputation.

We implement BSR by extending LMCache and vLLM with probe-only lookup, bounded
window retrieval, staging reuse, failed-block invalidation, HMA-aware
whole-prefix invalidation, and force-local recovery.  On DeepSeek-V4-Flash with
TP=2 across two DGX Spark nodes, one clean no-resume sweep grows external state
from 1.956 to 31.277\,GiB/rank while measured L1 RWS remains exactly
500.75\,MiB/rank at $W=32$, yielding a 63.959$\times$ largest-state
external-to-live-staging ratio.  A second fresh 524K-token run repeats the
largest-state acceptance result.  Four accepted window settings match the
source-accounted law $M_{\rm L1}=W\times16{,}408{,}576$ bytes/rank with zero
measured residual.  A 524K-token MTP-on state with 170 registered tensors
satisfies the same operational acceptance criterion, and evaluated tier and
rank-asymmetric failures expose either complete reuse or zero external reuse
before fallback.  Finally, a matched SSD optimization reduces 512K restore
TTFT from 43.1 to 17.6 seconds without changing RWS, showing that restoration
capacity and restore throughput can be controlled separately.
\end{abstract}
\section{Introduction}

Long-context LLM serving increasingly treats prior computation as reusable
execution state.  Hierarchical and disaggregated KV-cache systems retain this
state in host memory, SSD, or remote memory so that future requests can avoid
repeating long prefills \cite{cachedattention2024,mooncake2025,lmcache2025,strata2026}.
These systems enlarge \emph{retention capacity}: the amount of reusable state
that can exist outside the engine.  They do not by themselves determine the
local capacity required to restore that state.

Our upstream baseline makes the distinction measurable.  In the pinned LMCache
whole-plan path, a 32K prefix retains 1.956\,GiB/rank and first achieves full
reuse at a 2\,GiB L1 rung; 128K retains 7.823\,GiB/rank and first succeeds at
8\,GiB; 256K retains 15.646\,GiB/rank and first succeeds at 16\,GiB.  The
successful measured L1 peaks are 1.956, 7.824, and 15.648\,GiB/rank,
respectively.  External storage has preserved the state, yet the restore path
still needs local staging on the scale of the complete selected plan.

We call this transient requirement the \emph{restoration working set} (RWS).
RWS is distinct from the external state footprint and from the destination
memory needed for normal model execution.  It is the peak amount of local
staging simultaneously live while external state is converted back into
executable state.  This paper asks:

\begin{quote}
Can a serving runtime know the complete reusable external state, restore it
through an explicitly bounded local working set, and still preserve
request-level correctness and fail-closed recovery for heterogeneous
Tensor-Parallel state?
\end{quote}

\begin{figure}[H]
\centering
\begin{minipage}[t]{0.47\textwidth}
\centering
\textbf{(a) Whole-plan restoration}\par\vspace{2mm}
\begin{tikzpicture}[
  every node/.style={font=\small,align=center},
  box/.style={draw,rounded corners=2pt,minimum width=5.6cm,minimum height=8mm,inner sep=2mm},
  arr/.style={-{Latex[length=2mm]},thick},node distance=7mm]
  \node[box] (ext) {external reusable state $S$};
  \node[box,very thick,below=of ext] (stage) {materialize complete selected plan};
  \draw[arr] (ext) -- node[right,font=\footnotesize]{restore} (stage);
  \node[below=4mm of stage,font=\small] {$\operatorname{RWS}_{\rm whole}(S)=\Theta(|S|)$};
\end{tikzpicture}
\end{minipage}\hfill
\begin{minipage}[t]{0.47\textwidth}
\centering
\textbf{(b) Bounded-State Restoration}\par
{\footnotesize\bfseries complete knowledge, bounded residency}\par\vspace{1mm}
\begin{tikzpicture}[
  every node/.style={font=\small,align=center},
  box/.style={draw,rounded corners=2pt,minimum width=5.6cm,minimum height=8mm,inner sep=2mm},
  smallbox/.style={draw,very thick,rounded corners=2pt,minimum width=3.5cm,minimum height=8mm,inner sep=2mm},
  arr/.style={-{Latex[length=2mm]},thick},node distance=6mm]
  \node[box] (ext) {external reusable state $S$};
  \node[box,below=of ext] (probe) {probe complete reusable prefix\\without whole-plan staging};
  \node[smallbox,below=of probe] (win) {install one $W$-window};
  \draw[arr] (ext) -- (probe);
  \draw[arr] (probe) -- (win);
  \node[below=3mm of win,font=\footnotesize] {release staging; reuse for next window};
  \node[below=9mm of win,font=\small] {$\operatorname{RWS}_{\rm BSR}(S,W)=O(W)$};
\end{tikzpicture}
\end{minipage}
\vspace{2mm}

\fbox{\strut $|S|\uparrow\;\nRightarrow\;\operatorname{RWS}\uparrow\quad\text{for fixed }W$}
\caption{The abstraction change.  BSR separates complete hit discovery from
local residency and reuses one bounded staging window across the restore plan.}
\label{fig:novelty}
\end{figure}
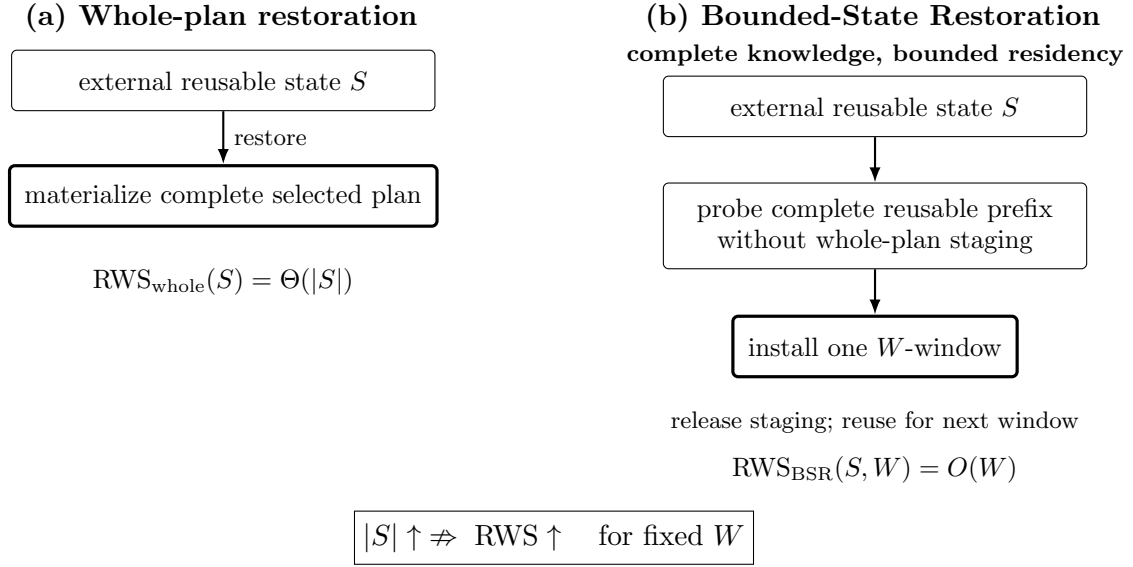

The central observation is that \emph{complete knowledge does not require
complete local residency}.  Figure~\ref{fig:novelty} contrasts the two resource
models.  A whole-plan path may couple peak restore staging to the complete
selected state.  \bsr separates the decision from materialization: the runtime
first discovers the complete reusable prefix without staging the whole hit, then
installs confirmed state through a reusable window of at most $W$ chunks.
Larger external state creates more iterations through the same local staging
budget rather than a larger simultaneously live restore set.

This is not merely smaller I/O.  The serving scheduler must know the complete
external hit before it can advertise computed tokens and allocate destination
blocks.  A lookup path that obtains that knowledge by materializing the same
complete hit has already paid the whole-plan space cost.  BSR therefore adds a
\emph{probe-only} phase whose output is knowledge---the complete candidate hit
set---rather than a whole-prefix local copy.

The second challenge is semantic.  Reusable execution state is not necessarily
a single homogeneous blob.  Our DeepSeek-V4-Flash configuration registers five
engine groups and eight kernel groups, with 167 tensors when MTP is disabled and
170 when MTP is enabled, and the state spans two TP ranks.  A partial transfer
can therefore deliver plausible bytes while leaving the advertised
request-level prefix unsafe.  BSR keeps semantic ownership with the runtime:
reuse is committed only if every required rank/group component is valid;
otherwise the advertised prefix is invalidated and the request falls back to a
lower valid tier or recomputation.

We implement this protocol by modifying LMCache and vLLM, without changing model
kernels or the external state representation.  The final evaluation is built
around two publication-grade capacity experiments: an upstream whole-plan L1
ladder and a clean same-epoch BSR scaling sweep.  Supporting experiments test
the $W$ memory law, heterogeneous MTP state, fail-closed recovery, and whether
restore throughput can improve without increasing RWS.

The main results are:
\begin{itemize}[leftmargin=1.5em]
\item Measured whole-plan full-reuse points first succeed at 2, 8, and
16\,GiB L1 for 1.956, 7.823, and 15.646\,GiB/rank external states; successful
L1 peaks track the complete selected plan.
\item In one clean same-epoch BSR sweep, external state grows from 1.956 to
31.277\,GiB/rank while measured L1 RWS stays exactly 500.75\,MiB/rank at
$W=32$, reaching a 63.959$\times$ external-to-live-staging ratio.  A second
fresh 524K run repeats the largest-state acceptance result.
\item At $W\in\{8,16,32,64\}$, measured RWS exactly equals
$W\times16{,}408{,}576$ bytes/rank.
\item A 524K-token MTP-on state with 170 registered tensors satisfies the same
operational acceptance criterion; evaluated tier and rank-asymmetric failures
expose either the complete advertised prefix or zero external reuse before
fallback.
\item Raising SSD object-load concurrency from 1 to 4 reduces 512K restore TTFT
from 43.1 to 17.6 seconds while measured RWS remains 500.75\,MiB/rank.
\end{itemize}

\section{Restoration Capacity as a Resource}
\label{sec:model}

Let reusable external execution state $S$ induce an ordered restore plan
$P(S)=\langle u_1,\ldots,u_N\rangle$ under a fixed model/runtime registration.
Restore unit $u_i$ occupies $c_i$ bytes in local staging while it is installed.
The exact unit is implementation-defined; in our implementation the control
parameter $W$ is expressed in LMCache chunks, and a chunk spans the registered
per-rank state required by that token range.

For restore execution $\rho$, define
\begin{equation}
\operatorname{RWS}(\rho)=\max_t\sum_{u_i\in L_\rho(t)} c_i,
\label{eq:rws}
\end{equation}
where $L_\rho(t)$ is the set of restore units simultaneously live in the local
staging tier at time $t$.  The definition excludes the externally retained copy
and the destination state required by normal model execution.  It measures the
transient local-capacity tax imposed by restoration itself.

This exposes three distinct resources.  \emph{Retention capacity} is how much
reusable state can exist outside the engine.  \emph{Restoration capacity} is the
RWS needed to make a selected state executable.  \emph{Execution headroom} is
what remains for model weights, active KV state, runtime buffers, and concurrent
work after the restore pool is provisioned.  An external tier primarily expands
the first quantity; the restore algorithm determines the second; provisioning
the second consumes the third.

\paragraph{Whole-plan materialization.}
If discovery constructs the complete selected plan and staging for all selected
units is reserved before those units can be released, then
\begin{equation}
\operatorname{RWS}_{\rm whole}(S)\ge\sum_{i=1}^{N}c_i.
\label{eq:whole}
\end{equation}
For a fixed layout with bounded positive unit sizes, this gives
$\Theta(N)=\Theta(|S|)$ restoration capacity.  We do not claim that every cache
system must have this property; the upstream ladder in Section~\ref{sec:eval}
measures it for the pinned whole-plan path used as our baseline.

\paragraph{Knowledge versus residency.}
Let $H(S)$ denote the complete candidate hit produced by external discovery.
The scheduler needs $H(S)$ before it can account for the externally computed
prefix, but the bytes represented by $H(S)$ need not be staged simultaneously.
BSR imposes the separate live-residency invariant
\begin{equation}
|L_\rho(t)|\le W\quad\forall t\text{ during bounded installation}.
\label{eq:live}
\end{equation}
Thus $H(S)$ is a decision contract; Eq.~\ref{eq:live} is a resource contract.

\paragraph{Bounded installation.}
BSR partitions the confirmed plan into ordered windows
$G_1,\ldots,G_K$, $|G_j|\le W$, with $K=\lceil N/W\rceil$.  A window's staging
is released before the same capacity is reused for the next window.  If at most
one window is live and auxiliary restore state does not allocate in proportion
to $N$, then
\begin{equation}
\operatorname{RWS}_{\rm BSR}(S,W)\le Wc_{\max}+M_{\rm aux}(W).
\label{eq:bounded}
\end{equation}

\textbf{Proposition 1 (capacity decoupling).}  Under the assumptions above,
for a fixed registration and fixed $W$, peak restore-pool capacity is $O(W)$
and independent of $N$.

\emph{Proof sketch.} At any time, live staged units belong to at most one
admitted window, whose cardinality is at most $W$.  Moving to the next window
changes the identity of live units, not the maximum simultaneous count.  Adding
the bounded auxiliary term yields Eq.~\ref{eq:bounded}.  $\square$

The proposition moves state growth from capacity growth to iteration growth:
$K=\lceil N/W\rceil$ increases with state size, while the live staging bound
remains fixed.  Total bytes still have to be moved and installed, so total work
remains $\Theta(|S|)$ and latency may grow with $|S|$.  BSR is a space
abstraction, not a constant-time restoration claim.

For the evaluated MTP-off DeepSeek-V4-Flash registration, source accounting
specializes Eq.~\ref{eq:bounded} to
\begin{equation}
\boxed{M_{\rm L1}(W)=W\times16{,}408{,}576\ \text{bytes/rank}.}
\label{eq:law}
\end{equation}
The coefficient is registration-specific; the portable abstraction is that $W$
controls live restoration capacity while $|S|$ controls the number of windows.

\section{Protocol: Probe, Install, Commit}
\label{sec:design}

A bounded staging loop is insufficient unless it composes with scheduling and
request semantics.  BSR therefore separates restoration into three phases:
\emph{probe}, \emph{install}, and \emph{commit}.

\begin{figure}[H]
\centering
\resizebox{0.98\textwidth}{!}{%
\begin{tikzpicture}[
  every node/.style={font=\small,align=center},
  box/.style={draw,rounded corners=2pt,minimum height=10mm,inner sep=2mm,text width=31mm},
  winbox/.style={draw,very thick,rounded corners=2pt,minimum height=10mm,inner sep=2mm,text width=35mm},
  okbox/.style={draw,rounded corners=2pt,minimum height=9mm,inner sep=2mm,text width=32mm},
  failbox/.style={draw,dashed,rounded corners=2pt,minimum height=9mm,inner sep=2mm,text width=38mm},
  arr/.style={-{Latex[length=2mm]},thick},note/.style={font=\footnotesize},node distance=9mm and 9mm]
  \node[box] (probe) {\textbf{Probe-only lookup}\\complete candidate hit; no whole-plan L1 residency};
  \node[box,right=of probe] (alloc) {Advertise prefix and allocate destination blocks};
  \node[winbox,right=of alloc] (win) {\textbf{Window $j$ ($\le W$)}\\L2 $\rightarrow$ L1 $\rightarrow$ GPU\\complete $\rightarrow$ sync $\rightarrow$ release};
  \node[box,right=of win] (gate) {\textbf{Semantic commit}\\all required ranks/groups valid?};
  \draw[arr] (probe)--(alloc); \draw[arr] (alloc)--(win); \draw[arr] (win)--(gate);
  \node[note,below=3mm of probe] {complete knowledge};
  \node[okbox,below=20mm of gate,xshift=7mm] (reuse) {$Q(S)=1$\\expose full-prefix reuse};
  \node[failbox,below=23mm of win,xshift=-7mm] (fail) {$Q(S)=0$ or window failure\\invalidate advertised prefix};
  \node[failbox,below=7mm of fail] (fallback) {lower valid tier or deterministic recompute};
  \draw[arr] (gate.south) -- node[right,note,fill=white,inner sep=1pt]{yes} (reuse.north);
  \draw[arr] (gate.south west) -- node[above left,note,fill=white,inner sep=1pt]{no} (fail.north east);
  \draw[arr,dashed] (win.south east) -- ++(0,-7mm) -| (fail.north);
  \draw[arr] (fail)--(fallback);
\end{tikzpicture}}
\caption{BSR is a request protocol, not merely smaller I/O.  Probe obtains the
complete candidate prefix without whole-plan local residency; install reuses a
$W$-bounded staging set; commit exposes reuse only if the complete advertised
request state remains valid.}
\label{fig:protocol}
\end{figure}
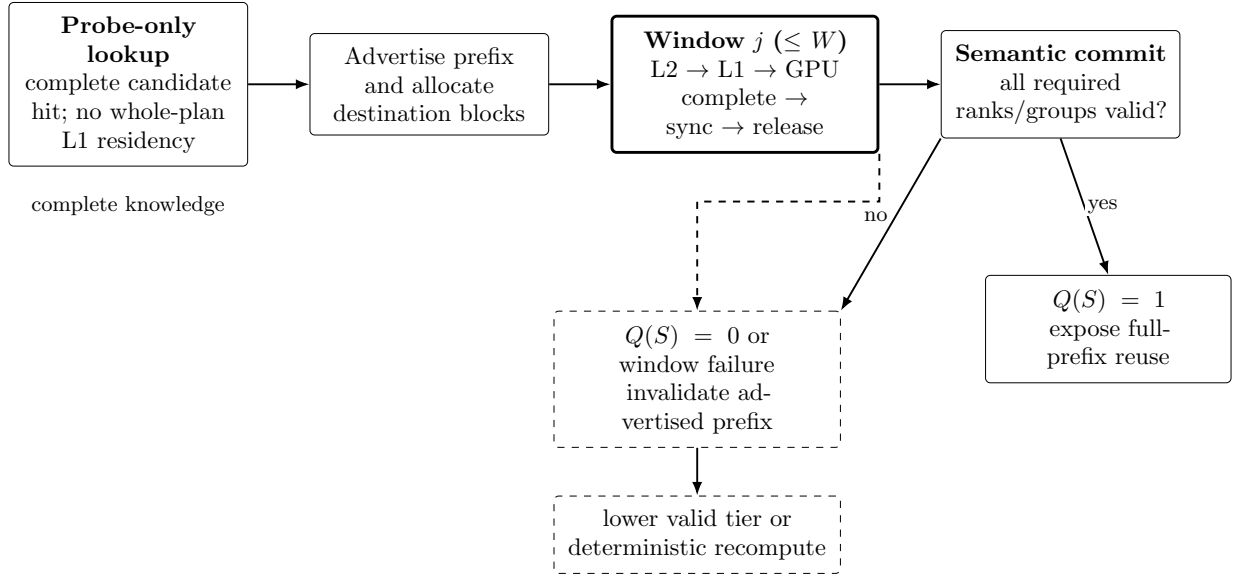

\paragraph{Probe globally.}
The scheduler needs a complete external-hit decision before allocating the
corresponding destination execution state.  Probe-only lookup asks the storage
layer to determine the retained hit set but deliberately avoids retaining the
complete L2 result in L1.  In particular, probe releases lookup-side L1/L2 read
locks and returns the hit bitmap.  The result is knowledge of what may be
restored, not residency of those bytes.

\paragraph{Install locally.}
After destination blocks are allocated, BSR iterates through the confirmed
prefix in windows of at most $W$ chunks.  Each window is fetched, required to be
complete, mapped onto the corresponding destination block slice, transferred,
synchronized, and then released.  Only after release may the staging storage be
reused by the next window.  The current prototype serializes bounded restores
with a dedicated lock so concurrent requests cannot multiply the configured
per-request staging budget.

\paragraph{Commit at request semantics.}
The storage layer knows objects; the runtime knows whether those objects compose
a valid request state.  Let $q_{r,g}$ indicate validity of the required state
for TP rank $r$ and semantic group $g$.  BSR commits the advertised prefix only
if
\begin{equation}
Q(S)=\bigwedge_{r=1}^{R}\bigwedge_{g=1}^{G}q_{r,g}=1.
\label{eq:q}
\end{equation}
If a window is incomplete, times out, or fails during transfer, the retrieve is
failed and its destination blocks are marked invalid.  For heterogeneous-memory
allocation, any invalid required group collapses the complete advertised
external prefix to zero computed tokens.  This conservative composition is
important because per-group block indices need not represent comparable token
offsets.

\paragraph{Recover without re-entering the same fault.}
A failed asynchronous restore must not immediately re-trigger the same external
lookup.  BSR therefore installs a force-local marker after failed remote
restoration.  The next admission pass reports zero external tokens, re-enters
normal local-prefill admission, and removes the marker only after successful
RUNNING admission or request teardown.  The resulting failure path is
request-level: lower valid tier when available, otherwise deterministic local
recomputation.

\textbf{Proposition 2 (fail-closed exposure under the protocol).}  If failed
window installation invalidates all affected destination blocks, HMA invalidity
collapses the advertised prefix, and reuse is exposed only when Eq.~\ref{eq:q}
holds, then no failed bounded restore is intentionally exposed as a partially
valid advertised prefix.  This is a protocol property under the stated
implementation rules; Section~\ref{sec:eval} evaluates it operationally rather
than claiming formal verification.

\section{Implementation}
\label{sec:impl}

We implemented BSR as a narrow control-path extension to pinned LMCache and
vLLM sources.  The external state representation and model kernels are
unchanged.  Table~\ref{tab:mechanisms} maps the implementation mechanisms to the
contracts above.

\begin{table}[t]
\centering
\caption{Implementation mechanisms.}
\label{tab:mechanisms}
\begin{tabularx}{\textwidth}{@{}p{0.27\textwidth}X@{}}
\toprule
Mechanism & Role \\
\midrule
Probe-only lookup & Return the complete retained hit while releasing lookup-side staging and read locks. \\
Restore-window configuration & Expose \texttt{prefetch\_window\_chunks}; zero retains legacy behavior. \\
Windowed retrieve & Fetch, verify, install, synchronize, release, and reuse one window at a time. \\
Bounded-retrieve lock & Prevent concurrent bounded requests from multiplying the configured live staging budget. \\
Failed-block invalidation & Ensure a failed partially installed retrieve cannot remain valid destination state. \\
HMA-aware invalidation & Collapse any invalid required heterogeneous group to whole-prefix invalidation. \\
Force-local recovery & Skip repeated external lookup after a failed restore and re-enter local-prefill admission. \\
FS load concurrency & Parallelize object reads inside a fixed window without changing $W$ or RWS. \\
\bottomrule
\end{tabularx}
\end{table}

\paragraph{LMCache changes.}
The distributed request specification gains a \texttt{probe\_only} mode and the
storage configuration gains \texttt{prefetch\_window\_chunks}.  Probe-only
lookup performs normal hit discovery but releases retained L1/L2 locks instead
of materializing the whole retained set.  The retrieve path branches on the
window setting: legacy behavior is preserved at zero; bounded mode invokes a
windowed routine.  For each admitted range, that routine submits a prefetch,
waits for complete presence, obtains the staged objects, slices the destination
block IDs to the same chunk range, performs H2D installation, synchronizes the
stream, releases the prefetched keys, and advances to the next range.  Any
missing key, timeout, or transfer exception fails the retrieve.

\paragraph{vLLM failure integration.}
A failed retrieve can already have copied a subset of destination blocks, so the
connector records all destination block IDs as erroneous on failure.  The vLLM
scheduler's HMA invalid-block path was extended to inspect all block groups.  If
any group contains an invalid block, the complete externally advertised prefix
is invalidated rather than trying to infer a shared token boundary from
heterogeneous block indices.  The force-local marker then prevents the next
scheduler pass from repeating the same external fault.

\paragraph{Storage-side concurrency is orthogonal.}
The filesystem adapter originally loaded objects serially.  We added a separate
\texttt{LMCACHE\_FS\_LOAD\_CONCURRENCY} knob implemented with an asynchronous
semaphore.  Reads fill the same preallocated L1 window objects; hit bitmap and
per-object success semantics are unchanged.  Thus this optimization can change
throughput without changing the restoration-space contract.

\paragraph{System topology.}
The evaluation uses DeepSeek-V4-Flash on two DGX Spark nodes with TP=2.  The
MTP-off registration contains 167 tensors and the MTP-on registration 170; the
LMCache chunk size is 256 tokens and the MTP-off $W=32$ window accounts to
16,408,576 bytes per chunk across registered state, or 500.75\,MiB/rank per
window.  Each DGX rank has a qualified direct 100\,Gb/s RDMA path to one
Mooncake store on an external host; the store is configured with a 128\,GiB
segment and strict tenant isolation.  Local SSD remains available as a lower
reusable tier.

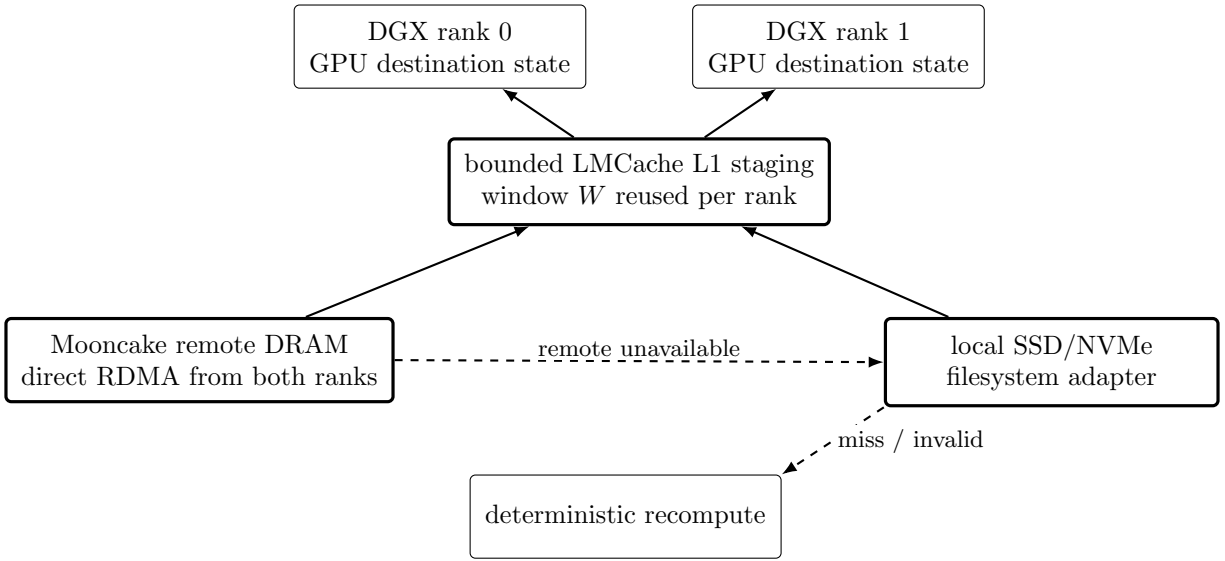
\begin{figure}[H]
\centering
\begin{tikzpicture}[
 every node/.style={font=\small,align=center},
 box/.style={draw,rounded corners=2pt,minimum width=37mm,minimum height=11mm,inner sep=2mm},
 tier/.style={draw,very thick,rounded corners=2pt,minimum width=44mm,minimum height=11mm,inner sep=2mm},
 arr/.style={-{Latex[length=2mm]},thick},node distance=12mm and 14mm]
 \node[box] (r0) {DGX rank 0\\GPU destination state};
 \node[box,right=of r0] (r1) {DGX rank 1\\GPU destination state};
 \node[tier,below=of $(r0)!0.5!(r1)$] (l1) {bounded LMCache L1 staging\\window $W$ reused per rank};
 \node[tier,below left=of l1,xshift=7mm] (mc) {Mooncake remote DRAM\\direct RDMA from both ranks};
 \node[tier,below right=of l1,xshift=-7mm] (ssd) {local SSD/NVMe\\filesystem adapter};
 \node[box,below=15mm of $(mc)!0.5!(ssd)$] (rec) {deterministic recompute};
 \draw[arr] (mc)--(l1); \draw[arr] (ssd)--(l1);
 \draw[arr] (l1)--(r0); \draw[arr] (l1)--(r1);
 \draw[arr,dashed] (mc.east)--node[midway,above,font=\footnotesize,fill=white,inner sep=1.5pt]{remote unavailable}(ssd.west);
 \draw[arr,dashed] (ssd.south west)--node[midway,right,font=\footnotesize,fill=white,inner sep=1.5pt]{miss / invalid}(rec.north east);
\end{tikzpicture}
\caption{Evaluated hierarchy.  Both TP ranks have qualified direct RDMA paths
to the same Mooncake store.  BSR bounds the transient L1 staging used while
state moves from a reusable tier into the two destination ranks.}
\label{fig:system}
\end{figure}

\section{Experimental Methodology}
\label{sec:method}

The evaluation is organized around two primary experiment blocks and three
supporting requalifications.  We use committed run artifacts as the measurement
source; no resumed or interrupted attempt contributes a row to the main
fixed-$W$ scaling result.  A successful restore is accepted only when the
expected prefix is reported cached, the intended tier counters are observed,
the canonical output hash matches cold execution, the registered state geometry
matches the experiment, and both TP ranks remain healthy.  Failure experiments
retain the same output and TP-health checks while cached-token accounting
distinguishes lower-tier reuse from recomputation.

\paragraph{Experiment 1: upstream whole-plan capacity ladder.}
We disable bounded prefetching ($W=0$) and increase configured L1 capacity
through discrete GiB rungs until full-prefix reuse succeeds.  The first
successful rung is therefore an empirical capacity requirement of the pinned
whole-plan path for that prefix.  The accepted ladder measures 32K at 2\,GiB,
128K at 8\,GiB, and 256K at 16\,GiB.  The 256K point is from a clean redo that
reports full reuse, the canonical output hash, and a measured L1 peak of
16,802,381,824 bytes/rank.  A prior engine-unhealthy 256K attempt is retained as
historical evidence but superseded by the clean PASS.  We did not complete a
512K/32\,GiB upstream run, so no 512K whole-plan measurement is used in the main
comparison.

\paragraph{Experiment 2: clean same-epoch BSR scaling.}
The primary bounded-space result is one no-resume E1 sweep
(\texttt{br-e1-0c12f749f250}) with $W=32$, MTP disabled, L1 configured at
1\,GiB, and direct RDMA transport.  It evaluates 32768, 65536, 131072, 262144,
393216, and 524032 tokens on the same deployed engines.  No engine relaunch
occurs within the sweep.  All six rows in the main scaling figure come from
this single experiment block.  A second fresh 524032-token run
(\texttt{br-e1-e956c433d35d}) uses a new salt on the same engines and repeats
the largest-state acceptance result.  Attempts interrupted before evidence
emission are excluded from the measured dataset.

\paragraph{Window accounting.}
Accepted $W\in\{8,16,32,64\}$ measurements at 256K are used only to test the
$W\rightarrow$RWS capacity law.  They were not collected as a randomized,
repeated latency frontier, so their TTFT values are descriptive rather than a
basis for a universal latency-optimal $W$.

\paragraph{Heterogeneous and failure requalification.}
A 524032-token MTP-on run requalifies the protocol with 170 registered tensors.
The failure matrix independently exercises remote-store unavailability,
remote-plus-SSD absence, and a one-rank remote restore fault.  We compare the
categorical cached-prefix outcome and acceptance checks, not failure TTFTs.

\paragraph{Performance A/B.}
A matched 512K SSD experiment holds model registration and $W=32$ fixed while
filesystem object-load concurrency changes from 1 to 4.  This experiment is
used only to test whether throughput can improve without increasing RWS.

\paragraph{Transport sanity.}
Isolated 4\,GiB Mooncake RDMA measurements reach 10.31 and 10.34\,GB/s GET and
10.83\,GB/s PUT per rank.  A synchronized concurrent rerun measures
18.87\,GB/s aggregate GET and 21.21\,GB/s aggregate PUT across the two ranks.
These values establish that the remote substrate is capable of high bandwidth;
they are not used as an exact request-level restore ceiling.

\section{Evaluation}
\label{sec:eval}

The evaluation asks five questions: (1) does the upstream whole-plan path
couple local restoration capacity to selected state size? (2) does BSR remove
that coupling in a clean long-context scaling block? (3) is $W$ a predictable
capacity control? (4) does the protocol remain safe for heterogeneous state and
failures? and (5) can restore throughput improve without increasing the RWS?

\subsection{Whole-plan restoration consumes whole-plan capacity}

The upstream ladder provides a direct baseline.  A 32K prefix retains
2,099,978,240 bytes/rank (1.956\,GiB) and first reaches full reuse at a 2\,GiB
L1 rung, where the measured L1 peak is 2,100,297,728 bytes/rank
(1.956\,GiB).  At 128K, 8,399,912,960 bytes/rank (7.823\,GiB) first succeeds
at 8\,GiB, with an 8,401,190,912-byte peak (7.824\,GiB).  At 256K,
16,799,825,920 bytes/rank (15.646\,GiB) first succeeds at 16\,GiB, with a
16,802,381,824-byte peak (15.648\,GiB).  Thus, at every measured successful
point, peak local restore usage tracks the complete selected state to within
small bookkeeping overhead.

\begin{figure}[H]
\centering
\includegraphics[width=0.88\textwidth]{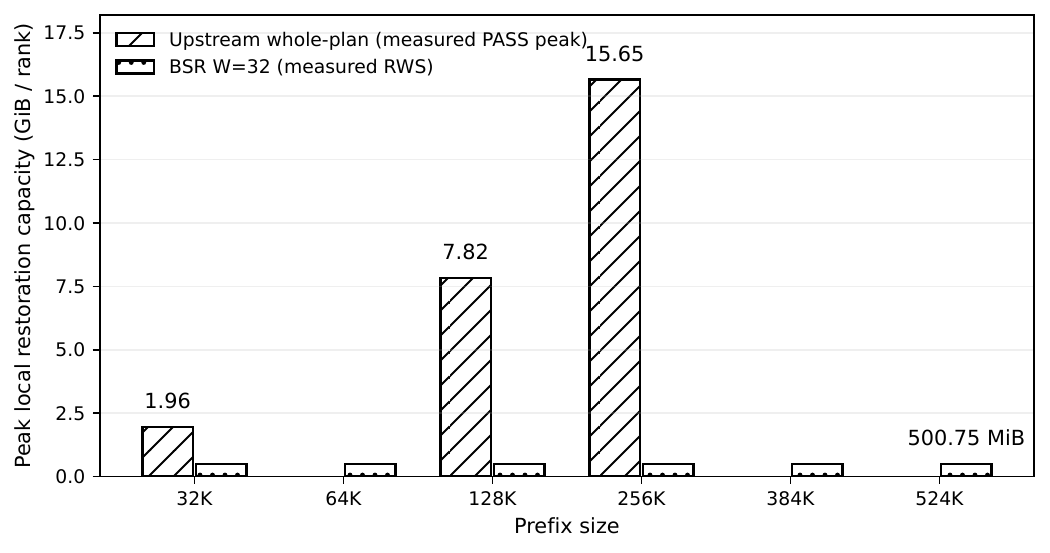}
\caption{Measured local restoration capacity.  Upstream whole-plan PASS peaks
rise with selected external state at 32K, 128K, and 256K.  BSR keeps the
measured $W=32$ RWS at 500.75\,MiB/rank across the clean scaling block.  No
512K upstream measurement is plotted.}
\label{fig:upstream}
\end{figure}

Figure~\ref{fig:upstream} compares those measured whole-plan PASS points with
BSR.  The figure intentionally contains no extrapolated 512K upstream bar: a
512K whole-plan run was not completed and is not treated as a measurement.

This is the practical distinction between external capacity and restoration
capacity.  Adding external storage can preserve a larger reusable state while
the whole-plan restore path still requires proportionally more local staging.
BSR converts that growth into more bounded-window iterations instead.

\subsection{A clean same-epoch sweep holds RWS fixed}

Experiment~2 grows external state from 2,099,978,240 to 33,583,245,760
bytes/rank, or from 1.956 to 31.277\,GiB/rank.  This is a 15.9921875$\times$
increase in retained state.  Across all six sizes, both TP ranks report exactly
525,074,432 bytes of peak L1 restore usage, i.e., 500.75\,MiB/rank.  All six
points report full cached-token reuse and the canonical output hash.  The
observed range of the measured L1 peak is therefore 0 bytes.

\begin{figure}[H]
\centering
\includegraphics[width=0.88\textwidth]{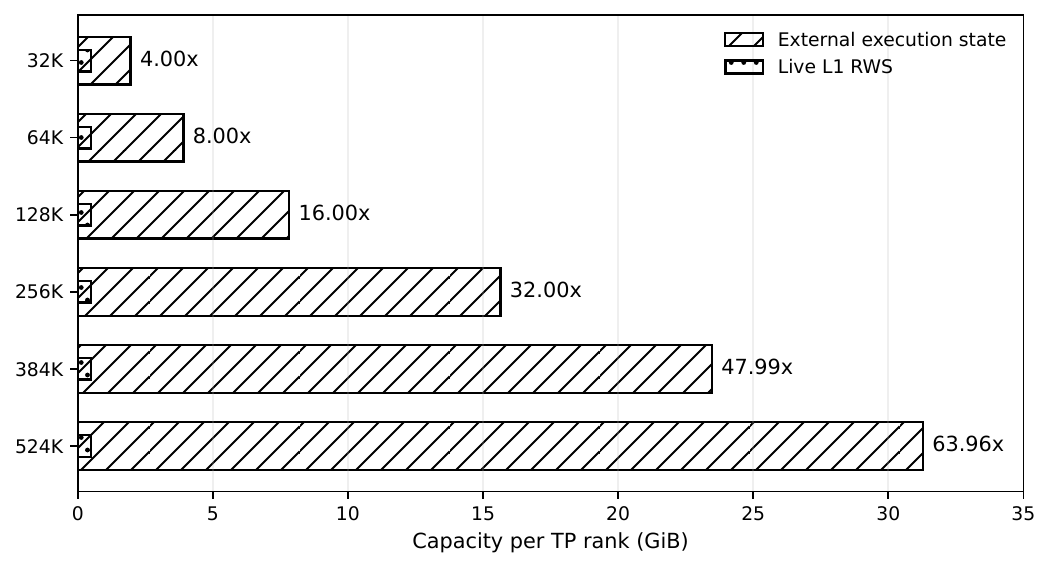}
\caption{Clean same-epoch fixed-$W=32$ scaling.  All six rows come from one
no-resume sweep on the same engines.  External state grows to 31.277\,GiB/rank
while measured L1 RWS remains 500.75\,MiB/rank.}
\label{fig:clean-scaling}
\end{figure}

At 524032 tokens, the external state is 63.9590$\times$ larger than the live L1
staging set.  Figure~\ref{fig:clean-scaling} visualizes the two capacities on
the same axis.

The corresponding Mooncake restore TTFTs are 2.17, 2.20, 4.68, 9.95, 12.90,
and 16.41 seconds from 32K through 524K.  They characterize this run; they do
not imply constant-time restoration.  A second fresh 524032-token run with a
new salt also satisfies full-prefix reuse and the same acceptance contract.
The capacity conclusion does not depend on combining results across epochs.

\subsection{\texorpdfstring{$W$}{W} is an explicit capacity control}

At 256K tokens, accepted settings $W=8,16,32,64$ measure L1 peaks of
125.1875, 250.3750, 500.7500, and 1001.5000\,MiB/rank.  Every point exactly
matches Eq.~\ref{eq:law}; the maximum residual from the source-accounted law is
0 bytes.

\begin{figure}[H]
\centering
\includegraphics[width=0.68\textwidth]{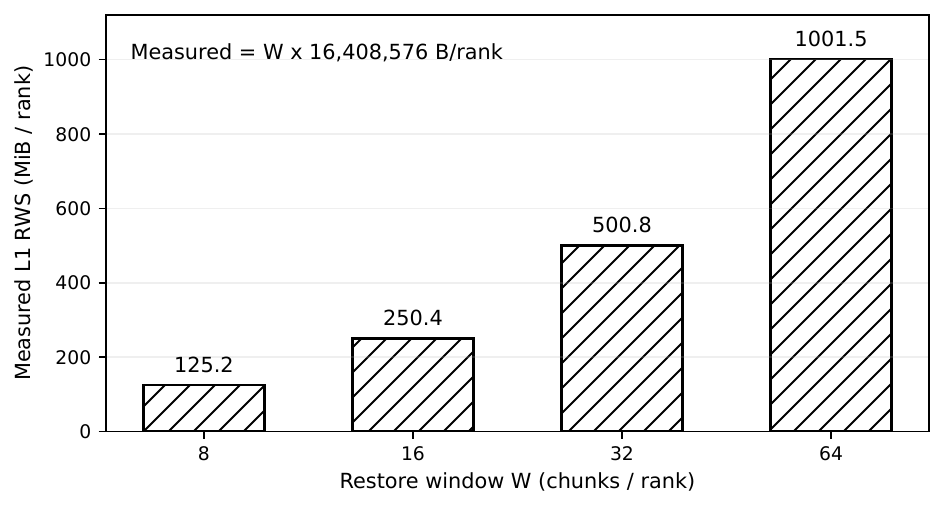}
\caption{Measured RWS as a configuration-controlled resource.  The four
accepted settings exactly match $M_{\rm L1}=W\times16{,}408{,}576$
bytes/rank.}
\label{fig:window}
\end{figure}

The same points have non-monotonic single-run Mooncake TTFTs of 7.54, 4.13,
8.09, and 7.05 seconds.  Because this block was collected for memory
accounting rather than as a controlled repeated performance frontier, we draw
no causal $W\rightarrow$latency conclusion and do not select a universal
latency-optimal $W$.

\subsection{Heterogeneous state and failures remain fail-closed}

The hardest accepted heterogeneous case enables MTP at 524032 tokens.  The
runtime registers 170 tensors; external state is 33,583,245,760 bytes/rank;
measured L1 peak is 532,283,392 bytes/rank (507.625\,MiB); the remote-only
restore reports all 524032 tokens cached, zero filesystem load, the canonical
output hash, and healthy TP=2 execution.  We treat this as a standalone
semantic requalification rather than a causal MTP-on/off comparison.

Figure~\ref{fig:fail} summarizes request-level failure outcomes.  With remote
storage unavailable and SSD valid, the complete 262144-token cached prefix is
reused from SSD.  With both reusable tiers absent, cached tokens are zero before
local recomputation.  A one-rank remote restore fault after one window likewise
collapses to zero external reuse and whole-prefix recomputation.  Every
evaluated case matches the canonical output and leaves both TP ranks healthy.

\begin{figure}[H]
\centering
\begin{tabularx}{0.96\textwidth}{@{}X>{\centering\arraybackslash}p{0.18\textwidth}>{\centering\arraybackslash}p{0.15\textwidth}>{\centering\arraybackslash}p{0.12\textwidth}@{}}
\toprule
Condition & Advertised cached prefix & Canonical output & TP=2 health \\
\midrule
Remote healthy / qualified restore & full & pass & pass \\
Remote unavailable, SSD healthy & full & pass & pass \\
Remote + SSD absent & zero & pass & pass \\
One-rank remote restore fault & zero & pass & pass \\
\bottomrule
\end{tabularx}
\vspace{1mm}
\caption{Observed request-level outcomes.  Across the evaluated failures,
reusable state is exposed as the complete advertised prefix or as zero external
reuse before fallback; no partial advertised reuse is observed.}
\label{fig:fail}
\end{figure}

\subsection{Throughput is separable from the RWS bound}

In the matched 512K SSD A/B, only filesystem load concurrency changes.  Raising
it from 1 to 4 reduces restore TTFT from 43.1 to 17.6 seconds, a 2.45$\times$
speedup, while measured L1 RWS remains exactly 525,074,432 bytes/rank in both
runs.  Server-side per-window load falls from roughly 0.52 seconds to 0.118
seconds.  The result demonstrates that storage-side throughput can improve
without expanding the restoration working set.

\begin{figure}[H]
\centering
\includegraphics[width=0.62\textwidth]{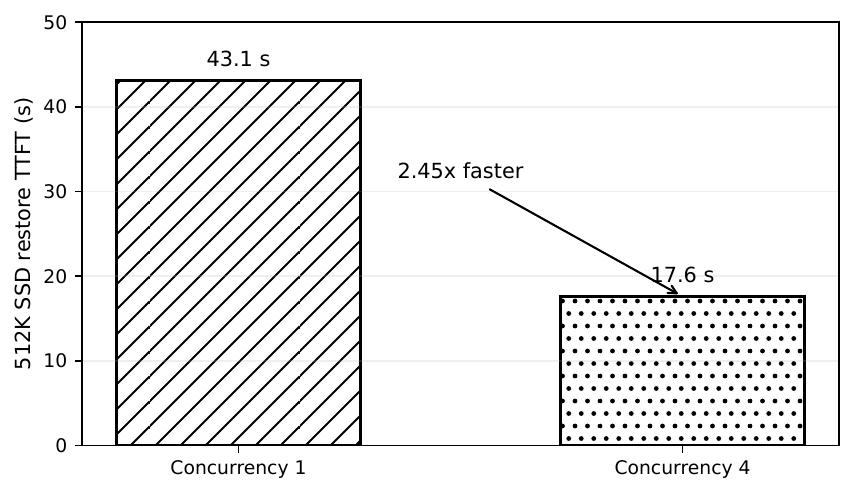}
\caption{Matched 512K SSD A/B at fixed $W=32$.  Storage-side concurrency
improves TTFT by 2.45$\times$ while measured RWS remains 500.75\,MiB/rank.}
\label{fig:perf}
\end{figure}

This performance result is deliberately secondary.  BSR's primary contribution
is the capacity contract; storage scheduling, transport, and installation
optimizations remain independent opportunities within that contract.

\section{Discussion and Limitations}
\label{sec:discussion}

\paragraph{RWS is not total process memory.}
The strongest invariant measured here is the LMCache L1 restore pool.  BSR does
not claim constant whole-process UMA: model weights, destination KV state,
allocator metadata, communication buffers, and unrelated runtime state can all
vary.  The abstraction is useful precisely because it isolates one controllable
transient capacity from those other resources.

\paragraph{The coefficient is not universal.}
The 16,408,576-byte per-$W$ coefficient is specific to the evaluated
DeepSeek-V4-Flash MTP-off registration and chunk geometry.  Another model or
state representation may have a different $c_i$.  The general requirement is a
bounded per-window footprint and auxiliary state that does not grow with the
full restore plan.

\paragraph{Current concurrency is conservative.}
The prototype serializes bounded restores globally with a dedicated lock.  This
makes the memory contract easy to audit but limits multi-request overlap.  A
natural extension is a global admission budget $B$ that allows several request
windows concurrently whenever their aggregate RWS remains below $B$.  That
would preserve the abstraction while improving service-level parallelism.

\paragraph{The 512K upstream point is derived, not measured.}
The whole-plan ladder has measured PASS points through 256K.  A clean
512K/32\,GiB whole-plan run was not completed.  We therefore omit 512K from
the measured baseline figure and do not claim a 512K upstream failure or
success.

\paragraph{Correctness is operational.}
The acceptance criterion combines cached-token accounting, tier isolation,
registration identity, canonical output equality, and post-request TP health.
It is substantially stronger than object-count or byte-count agreement, but it
is not a formal proof that every hidden state value is equivalent for every
possible prompt and future model.  Likewise, the fail-closed proposition assumes
the invalidation and recovery mechanisms execute as implemented.

\paragraph{BSR occupies one point in a broader design space.}
Direct coherent-memory systems can eliminate staging entirely on suitable
hardware \cite{directkv2026}; other restoration systems optimize scheduling,
layout, or recomputation/I/O overlap \cite{hcache2025,strata2026,cacheflow2026}.
BSR targets the common case where reusable state resides in an external tier
and must be staged and installed locally.  Its contribution is to make that
staging capacity explicit and bounded, then compose the bound with request
semantics.

\paragraph{Performance remains open.}
The clean E1 run shows restore latency growing with state size even while RWS is
flat, exactly as the model permits.  The filesystem A/B demonstrates one large
software optimization without changing RWS, but the remote path still contains
lookup, materialization, installation, and synchronization overheads.  We do
not claim that the current implementation is latency-optimal.

\section{Related Work}

\paragraph{Hierarchical and disaggregated KV retention.}
CachedAttention retains conversational KV state across a hierarchy and overlaps
loading with computation \cite{cachedattention2024}.  Mooncake builds a
KV-centric disaggregated serving architecture using CPU, DRAM, SSD, and network
resources \cite{mooncake2025}.  LMCache exposes KV cache as a reusable storage
and communication layer across inference engines and tiers \cite{lmcache2025}.
SYMPHONY disaggregates compute and KV memory and uses advisory prefetching plus
cooperative memory management \cite{symphony2026}.  These systems primarily
expand or orchestrate where reusable state resides.  BSR isolates the transient
local capacity required after external retention has succeeded.

\paragraph{State-restoration performance.}
HCache restores state from intermediate activations and co-schedules compute and
I/O \cite{hcache2025}.  Strata addresses fragmented hierarchical KV I/O with
GPU-assisted transfers and cache-aware scheduling \cite{strata2026}.  CacheFlow
models restoration as parallel work across tokens, layers, and GPUs
\cite{cacheflow2026}.  These works target restore latency and resource overlap.
BSR is complementary: its primary object is peak live restore staging, and its
protocol separates complete hit discovery from bounded residency before
request-level commit.

\paragraph{Avoiding staging.}
DirectKV shows that on coherent CPU--GPU platforms such as GH200/GB200, kernels
can directly access CPU-resident KV state and eliminate a GPU staging buffer
\cite{directkv2026}.  This is an important neighboring design point: eliminating
staging is stronger when the hardware and kernels permit it.  BSR instead
addresses external tiers that still require local installation and asks how to
bound that installation's transient capacity.

We make no ``first'' claim.  Streaming and bounded buffers are established
systems techniques.  The contribution is the serving-runtime contract needed
to apply bounded materialization to externally reusable LLM execution state:
the complete reusable prefix must be known before admission; the state spans
model-owned heterogeneous groups and TP ranks; and partial installation must not
be exposed as a valid request prefix.

\section{Conclusion}

External memory answers where reusable LLM execution state can live.  BSR
separates that retention question from a second resource: how much local
staging must be live while the state becomes executable again.  The protocol
first obtains complete knowledge of the reusable prefix, then installs it
through a reusable $W$-bounded working set, and commits reuse only after
request-level semantic validity is established.

The final experiment closure makes the capacity distinction directly visible.
In the measured upstream whole-plan path, 1.956, 7.823, and 15.646\,GiB/rank
states first achieve full reuse at 2, 8, and 16\,GiB L1 rungs, and successful
L1 peaks track the complete selected plan.  In one clean same-epoch BSR sweep,
external state grows to 31.277\,GiB/rank while measured L1 RWS remains exactly
500.75\,MiB/rank at $W=32$.  The largest state is 63.959$\times$ the live
staging set.  The measured $W$ sweep matches source accounting with zero-byte
residual, a 524K MTP-on state satisfies the same operational acceptance
criterion, and evaluated tier/rank failures expose complete reuse or zero
external reuse before fallback.  A matched SSD optimization then improves
512K restore TTFT by 2.45$\times$ without changing RWS.

The broader systems lesson is that \emph{external capacity, restoration
capacity, execution headroom, and restore throughput are distinct resources}.
A serving runtime can retain a large reusable state without requiring that the
same state be simultaneously resident in the local restoration tier.

\appendix
\section{Evidence Snapshot and Acceptance Contract}
\label{app:evidence}

This manuscript is anchored to repository
\path{StarkLeeSunny/Flexkv-doublenode}, branch
\path{feature/lmcache-bounded-lzzx1}, evidence ref \texttt{92c17bc}.  Primary
capacity claims are taken from committed raw evidence and
\path{program/paper-bounded-state-dataset.csv}; the manuscript does not use
conversational estimates or interrupted pre-emission attempts as measurements.

\begin{table}[H]
\centering
\caption{Primary evidence used by this manuscript.}
\footnotesize\begin{tabularx}{\textwidth}{@{}p{0.30\textwidth}X@{}}
\toprule
Claim & Repository evidence \\
\midrule
Upstream 32K/128K capacity ladder & \path{upstream-large-l1-b9cd34e45061.json} \\
Upstream clean 256K@16 GiB PASS & \path{upstream-large-l1-06c1daa7a2a7.json} \\
Clean fixed-$W$ E1 sweep & run \path{br-e1-0c12f749f250} / committed E1 evidence \\
Second fresh 524K acceptance & run \path{br-e1-e956c433d35d} in closure notes \\
Paper numeric rows & \path{program/paper-bounded-state-dataset.csv} \\
Window law, MTP, failures, RDMA & \path{program/paper-bounded-state-graph.json} and referenced evidence \\
SSD concurrency A/B & \path{program/perf-analysis-restore-throughput.md} and referenced run evidence \\
Post-window production health & \path{program/prod-baseline-attestation-20260817-1436.json} \\
\bottomrule
\end{tabularx}
\end{table}

An accepted successful restore must: (i) bind to the intended model and state
registration; (ii) report the intended full cached-token prefix; (iii) exercise
the intended tier according to load counters; (iv) produce the canonical output
hash; and (v) leave both TP ranks healthy.  Failure cases use the same output
and health checks while cached-token accounting must match the intended
fallback: full lower-tier reuse or zero external reuse before local
recomputation.

\section{Measured Capacity Tables}

\begin{table}[H]
\centering
\caption{Upstream whole-plan measured PASS points.}
\begin{tabular}{rrrr}
\toprule
Prefix & External state/rank & First PASS L1 rung & PASS L1 peak/rank \\
\midrule
32K & 1.956 GiB & 2 GiB & 1.956 GiB \\
128K & 7.823 GiB & 8 GiB & 7.824 GiB \\
256K & 15.646 GiB & 16 GiB & 15.648 GiB \\
\bottomrule
\end{tabular}
\end{table}

No 512K upstream PASS/FAIL row is reported because no clean 32\,GiB whole-plan
measurement was completed.

\begin{table}[H]
\centering
\caption{Clean same-epoch BSR scaling, $W=32$, MTP off.}
\begin{tabular}{rrrrr}
\toprule
Tokens & External GiB/rank & Objects/rank & L1 RWS MiB/rank & Cached \\
\midrule
32768 & 1.956 & 128 & 500.75 & 32768 \\
65536 & 3.912 & 256 & 500.75 & 65536 \\
131072 & 7.823 & 512 & 500.75 & 131072 \\
262144 & 15.646 & 1024 & 500.75 & 262144 \\
393216 & 23.469 & 1536 & 500.75 & 393216 \\
524032 & 31.277 & 2047 & 500.75 & 524032 \\
\bottomrule
\end{tabular}
\end{table}

\section{Implementation Surface}

The BSR patch surface is intentionally narrow.  LMCache adds probe-only lookup,
a restore-window configuration, windowed retrieve/release logic, and failed
retrieve block invalidation.  vLLM adds HMA-aware whole-prefix invalidation and
a force-local retry marker.  A separate filesystem patch adds within-window
read concurrency.  Model kernels, the registered external state representation,
and the model's attention semantics are unchanged.

\section*{AI Assistance Disclosure}
This work used DeepSeek-V4-Flash-0731 as an execution agent for engineering
tasks and experiment execution, and ChatGPT (GPT-5.6 Sol, High reasoning) for
research-plan discussion, system and experimental design, and manuscript and
figure design.  The author defined the research questions, reviewed and
approved system changes and experiments, verified the underlying evidence and
reported measurements, and is responsible for all claims and conclusions.

\end{document}